\documentstyle[prd,preprint,aps,epsfig]{revtex}

\begin{document}

\baselineskip 1.2 \baselineskip

\newcommand{\TeV}{\,{\rm TeV}}
\newcommand{\GeV}{\,{\rm GeV}}
\newcommand{\MeV}{\,{\rm MeV}}
\newcommand{\keV}{\,{\rm keV}}
\newcommand{\eV}{\,{\rm eV}}
\def\ap{\approx}
\def\bea{\begin{eqnarray}}
\def\eea{\end{eqnarray}}
\def\bi{\begin{itemize}}
\def\ei{\end{itemize}}
\def\be{\begin{enumerate}}
\def\ee{\end{enumerate}}
\def\ler{\lesssim}
\def\gtr{\gtrsim}
\def\beq{\begin{equation}}
\def\eeq{\end{equation}}
\def\haf{\frac{1}{2}}
\def\nn{\nonumber}
\def\p{\prime}
\def\ccg{\cal G}
\def\L{\cal L}
\def\O{\cal O}
\def\R{\cal R}
\def\U{\cal U}
\def\V{\cal V}
\def\W{\cal W}
\def\e{\varepsilon}
\def\slash#1{#1\!\!\!\!\!/}

\setcounter{page}{1}
\draft \widetext \preprint{KIAS-P01028, hep-ph/0106072}

\title{\Large \bf To be (finite) or not to be, that is the question
\\``Kaluza-Klein contribution to the Higgs mass''}

\author{Hyung Do Kim}

\address{Korea Institute for Advanced Study\\
{\tt hdkim@kias.ac.kr}}

\date{\today}

\tighten

\maketitle

\begin{abstract}

Recently five dimensional supersymmetric models with a
Scherk-Schwarz supersymmetry breaking and a localized
superpotential on a fixed-point have been constructed to yield a
definite prediction for the Higgs mass. We examine this issue in
detail and show that the finite one loop correction
and the definite prediction for the Higgs mass are
just a consequence of a special 
``Kaluza-Klein regularization'' scheme.

\end{abstract}

\newpage

\section{Introduction}

Electroweak symmetry breaking is one of the important problem in
high energy physics and the clear understanding of it is still
missing. Radiative electroweak symmetry breaking due to the large
Yukawa coupling is one of the appealing aspects of the
conventional supersymmetric models, so called MSSM 
(Minimal Supersymmetric Standard Model).
Weak scale supersymmetry has been the most popular model 
beyond the Standard Model 
due to several attractive aspects 
including the stabilization of Higgs mass, the gauge coupling unification and 
good dark matter candidates (LSP in the MSSM), 
but the complete theory including 
the supersymmetry breaking sector 
and the messenger sector is quite complicated and does not look economic.
As an alternative to the weak scale supersymmetry, 
several new ideas \cite{ADD,RS} considering extra dimensions
got an interest for recent several years.
Combining these ideas allows us to have a Scherk-Schwarz 
supersymmetry breaking mechanism \cite{ADPQ,DPQ,DQ} once 
the compactification radius is of order $1/{\rm TeV}$.
Also new mechanisms for the generation of electroweak symmetry
breaking has been proposed in these models. 
In the model \cite{BHN} it has been
claimed that Higgs mass can be predicted once Higgs VEV is
determined from Z boson masses with the aid of the finite one loop
correction of Kaluza-Klein(KK) contributions to the Higgs mass. The
result is insensitive to ultraviolet(UV) physics and all the contributions
of energy scales higher than the compactification scale is exponentially
suppressed \cite{AHNSW,HN}.
There has been a suspect \cite{GN} whether the result is just an artifact of
the special regularization choice, ``Kaluza-Klein
regularization'', or it is independent of the regularization
choice. It has been shown in \cite{GN} that the result is highly
sensitive to the cutoff of the Kaluza-Klein modes. Consequent papers
\cite{DGJQ,CP,N,W} revealed the opposite conclusion that the UV
insensitive result is robust. 
The result of \cite{DGJQ} shows that the finite size
resolution of the brane Yukawa coupling makes the effective Yukawa
couplings of higher KK modes soften and allows us to have a UV
insensitive result.
In \cite{CP} it has been shown that
the finiteness of the KK one loop correction is maintained in all
different regularizations which keep the supersymmetry manifest.
In \cite{N,W} five dimensional mixed position-momentum space has been
considered to show the ultraviolet(UV) insensitive characters
of the physics.
In this paper we follow the calculations given in the papers
and show that the finiteness of the one loop correction is based
on the infinite sum of the Kaluza-Klein states which is unphysical.
Under very reasonable assumptions that the momentum cutoff should be
given isotropically in five dimensional theory
($p_0,p_i$ and $p_5 = m_{\rm KK}$),
the truncated series can not be approximated by the infinite sum.
It is shown that the contributions of heavier Kaluza-Klein modes
are not exponentially suppressed and are equally important
for given four dimensional momentum.
Thus it is important to see the contributions of higher four dimensional
momentum near the physical cutoff since these are the main contributions
in the quadratically divergent corrections.
For fixed four dimensional momentum near the cutoff, 
the integrand appearing in the Higgs
mass one loop corrections becomes $e^{-p}$ after summing up all
the Kaluza-Klein modes, in other words infinite $p_5$ integration.
However, this is not the right procedure
since the inclusion of the Kaluza-Klein modes whose masses are heavier
than the cutoff does not have any sense.
The exponential suppression of the four dimensional momentum
brings us the finite result even after sending
our four dimensional cutoff to infinity.
It is very unnatural that the conclusion can be obtained only
when we summed up all the Kaluza-Klein states first.
In the next section, we briefly review the basic setup.
Then one example of the series is illustrated which is directly
related to the Kaluza-Klein contributions and shows clearly
what gives wrong interpretations.
The argument based on mixed position-momentum space is reexamined
and it is shown that the supersymmetry breaking is in the bulk
and exponential suppression of heavier Kaluza-Klein modes are not true.
Using these, we comment on the papers claiming the UV insensitiveness
that all the results are based on the infinite sum of Kaluza-Klein states.
Since supersymmetry is broken even above the compactification scale
(actually supersymmetry is broken at the fundamental scale
in the Scherk-Schwarz mechanism).
supersymmetry is not a symmetry that we should keep in choosing
the proper regularization.
Finally we summarize it.

\section{basic setup}

The constrained standard model \cite{BHN} from extra dimension can
be summarized as follows.
\footnote{It is well summarized in \cite{N}, 
and we do not repeat the explanation in detail.}
Consider five dimensional theory with
supersymmetry whose minimal multiplet corresponds to N=2
multiplets in 4-D language. After the compactification of one
extra dimension to $\frac{S_1}{Z_2 \times Z^{\prime}_2}$, we
obtain the Kaluza-Klein spectrum whose zero modes are the same as
that of the conventional standard model. One $Z_2$ breaks half of
the supersymmetry and the other $Z^{\prime}_2$ breaks the other
supersymmetry. Among $N=2$ vector multiplets, only vector fields
can have zero modes under this orbifold compactification since
others carry nonzero parity under the discrete symmetry.
the quarks and leptons have zero modes
after the compactification by the same reason
from $N=2$ hypermultiplets. After all if we
look at the spectrum below the compactification scale ($1/R$), it
is the same as that of the standard model. In the standard model
one loop correction to the Higgs mass is known to be quadratically
divergent and it becomes one of the important theoretical
motivations towards new theories beyond the standard model.
However, in the constrained standard model with extra dimensions
it is claimed that the quadratic divergences do not appear and the
one loop correction is determined independently of the detailed
nature of the fundamental scale physics. In this paper we examine
the calculation based on physical grounds and show that the
finiteness is an artifact of the Kaluza-Klein regularization, and
the result of the calculation is highly sensitive to the UV
physics.

\section{series convergence}

The question of whether the series converges is the basic one that
can be asked first. Any truncation of the series gives quadratic
divergences except an infinite sum reminds us divergent series.
For instance, Taylor series expansion of \bea \frac{1}{1+x} & = &
1 - x + x^2 - x^3 + \cdots \eea is a convergent series when $|x| <
1$ but is not a convergent one otherwise. When we use a formula
for the sum of a series, it is valid only when the series itself
is convergent (whether absolutely or conditionally). Infinite sum
has no meaning for divergent series. The remaining thing is to
show that the Kaluza-Klein sum has a similar problem with the
above non-convergent series.

There are two equivalent ways of calculating the one loop
contributions of the Kaluza-Klein states to Higgs mass. One loop
effective potential calculation is easy and we can get the mass
from it by differentiating the potential by Higgs fields.

\bea V_{\rm 1 loop eff} (\phi) & = & \frac{1}{2} {\rm tr}
\sum_{k=-\infty}^{\infty} \int \frac{d^4p}{(2\pi)} \log
\frac{p^2+m_{Bk}^2(\phi)}{p^2+m_{Fk}^2(\phi)}, \\ m_{Fk}(\phi) & =
& \frac{2k}{R} + m_t (\phi), \\ m_{Bk}(\phi) & = & \frac{2k+1}{R}
+ m_t (\phi), \eea

where $m_{Bk}$ and $m_{Fk}$ are masses of Kaluza-Klein states of
bosons and fermions. The most important contribution is from large
top Yukawa coupling and we restrict them to stop and top
respectively. $\phi$ is a convenient expression for the Higgs
field, and $R$ is the compactification radius and $k$ is an
integer representing the quantized momentum along the fifth
dimension.

Direct calculation of the one loop Higgs mass corrections involves
a sum of different Kaluza-Klein contributions.

\bea m_{H}^2 &\propto& y_t^2 \sum_{k=-\infty}^{\infty} \int
\frac{d^4p}{(2\pi)} \left[ \frac{1}{p^2 + m_{Bk}^2} -
\frac{1}{p^2+m_{Fk}^2} \right] \nn \\ &\propto&
\sum_{k=-\infty}^{\infty} \int \frac{d^4p}{(2\pi)} a_k, \nn \\ a_k
&=& \frac{m_{Fk}^2 - m_{Bk}^2}{(p^2+m_{Bk}^2)(p^2+m_{Fk}^2)}. \nn
\eea

Since \bea a_k \sim \frac{1}{p^4} \nn \eea at large four momentum
(for $p
> m_{F,B}$), at least logarithmic divergences are inevitable. For
the cutoff $\Lambda_c$ of the theory in which $\Lambda_c \sim N_c
(\frac{1}{R})$, we naturally expect

\bea
m_H^2 \propto -N_c (\frac{1}{R})^2 \log \Lambda_c
\eea

There are two useful expressions for the infinite series.

\bea \sum_{k=-\infty}^{\infty} (-1)^k \frac{x^3}{x^2+k^2} & = &
\frac{\pi}{2} \frac{x^2}{\sinh \pi x}, \\
\sum_{k=-\infty}^{\infty} \frac{x^3}{x^2+k^2} & = & \frac{\pi}{2}
\frac{x^2}{\tanh \pi x}. \eea

By substituting $x$ to $p$ and $k$ to $m_k$ with appropriate
numerical coefficients, these two expressions are used to derive
the finite one loop correction to the Higgs mass.

Furthermore, there are one more nice expressions that should be
kept in mind for later purpose.

\bea \frac{1}{2} \frac{1}{\sinh \pi x} & = &
\label{eq:n}
\sum_{n=-\infty}^{\infty} e^{-(2n+1)\pi x} \\ (x > 0) \nn
\eea

The above expressions are the key ingredients in deriving the
finiteness of the one loop correction to the Higgs mass. The above
series is definitely a convergent series and passes the first
test. Is it enough that the series is a convergent one? To have an
intuition of what is going on, let us look at one example.

\bea e^{-x} & = & 1 - x + \frac{x^2}{2!} - \frac{x^3}{3!} + \cdots
\eea

converges for any value of $x$. However, unless $x \ll 1$, first
several terms of the series show divergent behavior. For
concreteness, let us take $x=100$. The left-handed side gives
extremely tiny value $e^{-100} \sim 10^{-50}$, which is nearly
zero. On the other hand, the right-handed one is

\bea
1 - 100 + 5000 - \cdots
\eea

which looks like a divergent series with an amplifying magnitude
and alternating sign. The general profile is given 
in fig. 1 and fig. 2.
Furthermore, there are two ways of grouping the series. First,
summing the series with grouping two terms with ($+$,$-$) order
shows that

\bea e^{-x} & = & (1-x) + \frac{x^2}{2!} (1 - \frac{x}{3} ) +
\cdots + \frac{x^{98}}{98!} (1 - \frac{x}{99} ) \nn \\
&&+\frac{x^{100}}{100!} (1 - \frac{x}{101} ) + \cdots \eea

and the first 50 pairs give huge negative values for the series
with $x=100$. 
However, the result is very different for a different
pairing,
\bea e^{-x} & = & 1 - x(1-\frac{x}{2}) - \frac{x^3}{3!}
(1 - \frac{x}{4} ) + \cdots - \frac{x^{99}}{99!} (1 -
\frac{x}{100} ) \nn \\ && -\frac{x^{101}}{101!} (1 - \frac{x}{102}
) + \cdots \eea

which shows that the first 50 pairs give extremely huge positive
values with $x=100$. Generally the maximum of the envelop is at $N
= x$ ($N=100$ in the example with $x=100$ 
and $N=5$ for $x=5$ as in fig. 3). 
\begin{figure}
\vspace{5mm} \centerline{\epsfig{file=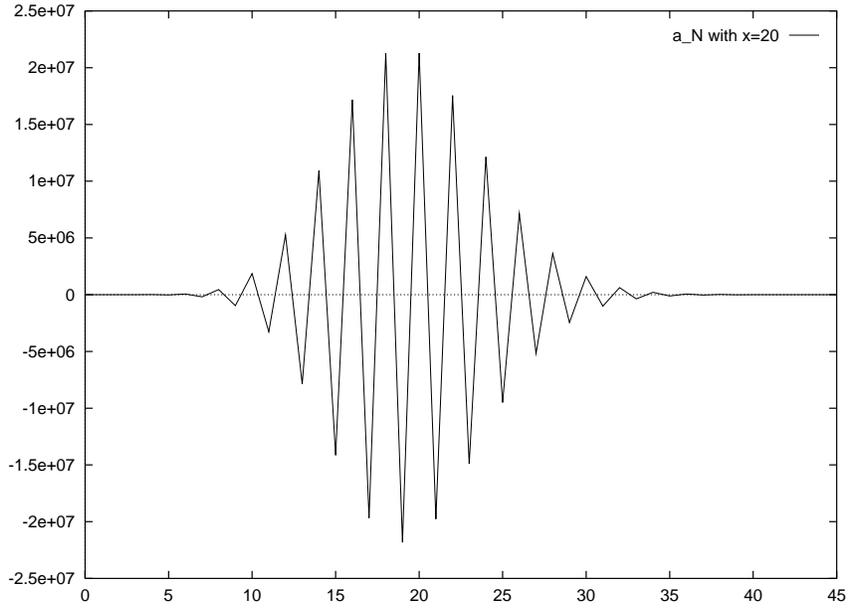, height=8cm}}
\label{series}
\vspace{5mm}
\caption{Divergent and convergent behavior of the truncated series for $x=20$.
$a_N = \sum_{k=0}^N (-1)^k \frac{x^k}{k!}$ and for $N=0$ to $N=45$.}
\end{figure}
\begin{figure}
\vspace{5mm} \centerline{\epsfig{file=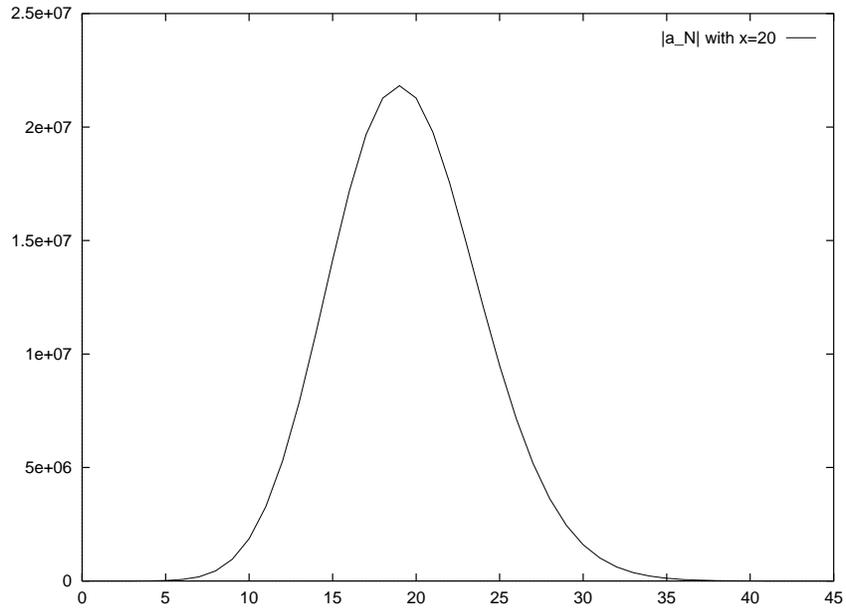, height=8cm}}
\label{seriesp}
\vspace{5mm}
\caption{Absolute value shows the magnitude of the truncated series for $x=20$.
For all regions $|a_N|$ is much larger than $e^{-x}$.}
\vspace{5mm}
\end{figure}

Fig. 3 shows at which values of $N$ the truncated series has a similar size
with the exact series ($e^{-x}$). For $x=5$, it is near $N \sim 15$.
Generally only beyond $N=ex$, 
the truncated series starts to show the convergent feature.

\begin{figure}
\vspace{5mm} 
\centerline{\epsfig{file=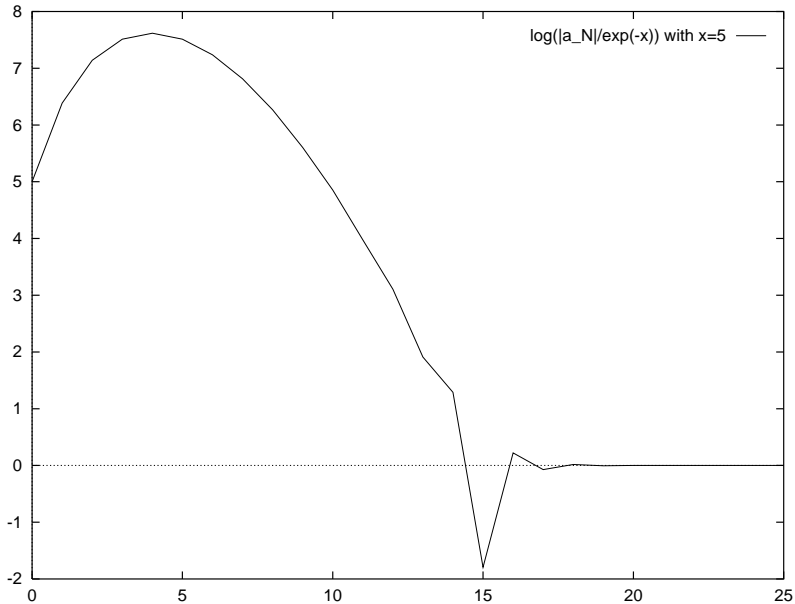, height=8cm}}
\label{ratio}
\vspace{5mm}
\caption{Logarithm of the ratio between $a_N$ and $e^{-x}$ for $x=5$.
For $N>5$, the ratio approaches to 1 (log approaches to 0), and only for $N>15$,
the ratio is close enough to 1 (log of the ratio is close enough to zero).}
\vspace{5mm}
\end{figure}

We can get a very important lesson
from this example. Even convergent series can show a strange
behavior for the truncation with finite sum. If the truncation
happens in the region that we can not see the convergent feature,
the approximation of replacing the sum over the finite modes to
the infinite sum is not valid anymore. Therefore to know which
point is the appropriate truncation capturing the underlying
physics is crucial in understanding the justification of the
Kaluza-Klein regularization scheme. Since we are dealing with
effective theories, there is a physical cutoff scale in the theory
beyond which our equipment (for example, perturbative quantum
field theory) is inappropriate. The cutoff is at or below the
fundamental scale in the five dimension which can be at most the
four dimensional Planck scale. Before reaching this fundamental
scale, we usually encounter the breakdown of the perturbation for
gauge couplings and Yukawa couplings. It is usually 10 or 100
times higher than the compactification scale due to the rapid
power running behavior in higher dimensions. Given the cutoff in
the theory, it is not physical to take into account all of the
Kaluza-Klein states which are heavier than the cutoff scale. From
the five dimensional theory point of view, Kaluza-Klein mass is
nothing but the momentum along the fifth dimension. The cutoff
$M_*$ apply to the momentum $p_0$, $p_i$ and $p_5$ equally in five
dimensions. This means that the momentum cutoff in the integral
should be comparable to the truncation of the Kaluza-Klein mass
terms which is $p_5$. If we accept this physical cutoff
regularization, the truncation point is located near the maximum
of the envelop of the series. In the example, the point is $k=x$
which shows a very divergent character. The cutoff of the four
momentum and Kaluza-Klein mass are highly correlated in the five
dimensional point of view, and the result based on the infinite
sum of Kaluza-Klein states is not justified for any value of the
four momentum cutoff. For all values of the cutoff $\Lambda_c$,
the series shows very unpredictable result which is highly
sensitive to the UV physics. The sign can not be fixed and this
fact reflects the quadratic divergent features directly. The
quadratic divergence means incalculability of the one loop, and
the usual $\Lambda_c^2$ contribution from the momentum cutoff in
analytically continued Euclidean space shows merely the order of
the corrections with undetermined sign.

A loop hole should be filled up in the above statements. It is the
validity of the momentum cutoff regularization. Usually good
regularization should preserve the symmetry that the original
theory has since we can not capture the physics correctly
otherwise. We are dealing with scalars and fermions, there is no
problem related to the gauge invariance. In \cite{DGJQ,CP,N,W} the
momentum cutoff regularization has been taken as inappropriate one
due to the fact that it does not keep the supersymmetry that the
original theory possesses. We show that there is no supersymmetry
remained after the Scherk-Schwarz compactification. This will be
discussed in detail in the next section.

\section{counter term}

In the paper \cite{BHN,AHNSW} the inability of writing down the
counter term for the Higgs mass has been used as one of the
supporting evidence for the finite result. If there is no way to
write down a counter term at tree level, this fact can be used as
a proof that the one loop calculation should be finite. It has
been claimed that the counter term can not be written due to
supersymmetry and discrete R parities. At one point $y=0$, one
supersymmetry is unbroken and the usual $\mu$ term is forbidden
due to the parity. ($H_1$ is even and $H_2$ is odd.) At other
point $y=\frac{\pi}{2} R$, the other supersymmetry survives and
also the $\mu$ term to different chiral multiplet ($H^{\prime}_1$
and $H^{\prime}_2$) is not allowed due to the different R parity.
There is a full N=2 supersymmetry in the bulk and the mass terms
for the hypermultiplets are forbidden also by R parity. Combining
these facts seems not to allow a mass term for the Higgs scalar,
$m_H^2 h^2$).

However, this argument is not precise and is based on several
misunderstandings of the setup. To understand how the
supersymmetry is broken in the extra dimension, it is helpful to
look at the wave functions of the component fields belonging to
the same supermultiplets. Fig. 4 shows one of the examples.

\begin{figure}
\vspace{5mm} \centerline{\epsfig{file=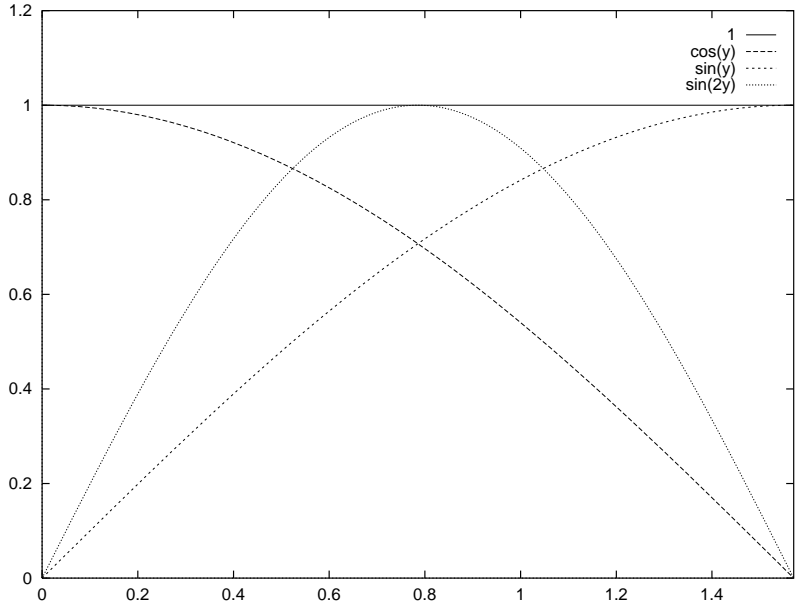, height=8cm}}
\label{wave}
\vspace{5mm}
\caption{Wave functions of first four modes in one $N=2$ multiplet
along the fifth dimension. Two of them form $N=1$ multiplet
at $y=0$ ($1$ and $\cos (y)$), and other two form $N'=1$ multiplet
at $y=\pi/2$ ($\sin (y)$ and $\sin (2y)$). ($R=1$ unit)
Away from $y=0$ and $y=\pi/2$, two wave functions are different
and supersymmetry can not be maintained anymore.}
\vspace{5mm}
\end{figure}

The restoration of one supersymmetry at the boundary is seen from
the agreement of two wave functions forming $N=1$ supermultiplets.
Just slightly away from the fixed point, we can see the difference
of two wave function profiles belonging to the same
supermultiplets and this can be interpreted as the breaking of the
supersymmetry in the bulk. Table 1 shows a qualitative difference
between the naive understanding of the supersymmetry breaking and
the actual situation.

\vspace{5mm}
\begin{tabular}{|c||c|c|c|c|}
\hline \hline
 & & $y=0$ & bulk & $y=\pi R/2$ \\
 & & (one fixed point) & & (the other fixed point) \\ \hline
SUSY (claimed in \cite{BHN,AHNSW}) & $N=1$ & broken & unbroken &
unbroken \\ \cline{2-5}
 & $N^{\prime} = 1$ & unbroken & unbroken & broken \\ \hline
 SUSY in real situation & $N=1$ & broken & broken &
unbroken \\ \cline{2-5}
 & $N^{\prime} = 1$ & unbroken & broken & broken \\ \hline \hline
\end{tabular}
\vspace{5mm}

If supersymmetry is broken and is mediated by heavy particles,
then it might be true that the contributions beyond the
compactification scale ($\frac{1}{R}$) is exponentially suppressed
and there may be a chance to get an answer which depends only
crucially on the physics at the compactification scale. However,
in the model considered here, the supersymmetry is broken in the
bulk and we start to feel it just away from the brane. Therefore,
the argument that forbids the counter term for the scalar Higgs
mass is not valid and $m_H^2 h^2$ is allowed in the bulk since the
supersymmetry is broken already. This indicates that the
Scherk-Schwarz SUSY breaking does not have a soft nature as you
might expect from the Kaluza-Klein regularization procedure.

\section{softness/hardness of Scherk-Schwarz SUSY breaking}

Scherk-Schwarz SUSY breaking uses different boundary conditions
for bosons and fermions along the compact extra dimension.
This condition has been imposed from the starting point
and the SUSY breaking scale is the fundamental scale
at which the theory is defined, even though the supersymmetry breaking
itself can be small enough by tuning the phase small
or making the compactification radius larger.
It is very different from the usual supersymmetry breaking mechanisms
in which supersymmetry is broken spontaneously at intermediate energy
scales which are relatively low compared to the fundamental scale.
For example, in gravity mediated one, SUSY breaking scale is of order
$10^{13}$ GeV, and in gauge mediation, it can vary from a few tens TeV
to intermediate scales.
Since the Scherk-Schwarz mechanism gives a hard breaking of supersymmetry,
there is no need to use a special regularization incorporating
supersymmetry in it.

The best way to look at the softness/hardness of Scherk-Schwarz SUSY
breaking is to investigate the wave functions of each KK modes in
the bulk. The direct five dimensional calculation might be a good
way to look at this problem. However, the indirect way of using KK
mode wave functions also shows the physics we want. Since the wave
functions belonging to the same $N=1$ supermultiplets are
different away from the fixed point, supersymmetry is broken
except the fixed point. This result is well summarized in the
table 1. 
This observation is very crucial in understanding the one
loop correction to the Higgs mass. It is argued that the
contribution from high momentum is exponentially suppressed
because the fluctuations happen in a small region near the fixed
point and can not feel the supersymmetry breaking effects located
far apart. This argument does not hold once supersymmetry is
broken just away from the fixed point. Even at high energies far
above the compactification scale, the fields involved in the loop
feel the supersymmetry breaking through the difference of the wave
functions in the bulk and these yield quadratic divergences in the
one loop calculation of the Higgs mass. Furthermore, the loop
integral is over the four momentum and this does not reflect the
physics discussed above. The momentum we are interested in is the
momentum along the fifth dimension $p_5$, in other words
Kaluza-Klein mass. If we carefully look at the contributions of
heavy Kaluza-Klein modes ($m_{k} \gg 1/R$), it is not
exponentially suppressed. This observation also shows that the
Scherk-Schwarz breaking of supersymmetry does not have a soft
nature. This is in accord with the observation of the bulk
supersymmetry breaking through the wave function difference
between the fields belonging to the same supermultiplet.

\section{other regularizations}

There exists a different viewpoint to the same problem. If the
starting theory is the ultimate theory and we know everything from
the start, then perhaps we should not truncate KK modes at any
arbitrary finite order. However, we know that 5-dimensional theory
also has a fundamental scale at which the gravity couples
strongly, and this prevents us from consideration of full
Kaluza-Klein modes. Thus we started from a theory whose limit is
clear. There is an upper bound for the cutoff which is below the
Planck scale. In this situation, to consider KK modes whose mass
is larger than the Planck scale is not the right thing that we
should do. If the series is convergent, then the answer would not
depend strongly on the truncation of KK modes and we can say
that our physical observable is really UV insensitive. However,
any truncation of infinite KK mode sum yields extremely strange
result and it shows that the result has a strong UV dependence.

The vanishing of tree level Higgs mass is now nothing but the
special choice of the counter terms which may not be the proper
one. Generic one loop correction overwhelms the tree level one,
and the choice of zero mass at the tree level is unnatural. In all
places of the extra dimensions they feel the SUSY breaking.
The SUSY breaking does
not reduce the degree of divergences for the scalar mass.
Therefore, Scherk-Schwarz SUSY breaking is the hard breaking and
quadratic divergences reappear. In this paper we have shown that
the Kaluza-Klein regularization is just one of special
mathematical regularization and other regularization gives
$y_t^2/16\pi^2 N_c (\frac{1}{R})^2$.

Now let us look at the papers supported the UV insensitiveness.
First, consider Gaussian spreading of the Yukawa couplings
\cite{DGJQ}.
Finite thickness $l_s = 1/\Lambda_s$ of the brane gives a field theoretic
 resolution of the apparent UV divergences.
For the branes with Gaussian distributions
(or equally for Yukawa couplings with Gaussian spreading),
\bea
f(y;l_s) & = & \frac{1}{\sqrt{2\pi} l_s} e^{-\frac{y^2}{2 l_s^2}}, \nn
\eea
the effective couplings of the nth KK modes are
\bea
y_t^n & = & y_t \exp \{ -\frac{1}{2} 
( \frac{M^n}{\Lambda_s}
)^2 \}. \nn
\eea
The above equation clearly shows that the suppression effects
due to the finite thickness of the brane appear at $\Lambda_s$
which is usually regarded as the fundamental scale.
This fact just tells us that the Kaluza-Klein modes heavier than
$\Lambda_s$ do not give an important contribution
to the Higgs mass correction.
Therefore, this method can be used as one of alternating procedure
to the finite $N$ truncation of Kaluza-Klein modes with momentum cutoff.
Unfortunately, this classical regularization still gives us misleading
answer if we sum up all the Kaluza-Klein modes.
The Poisson resummation formula is used in \cite{DGJQ} 
and it is essential in deriving the finite answer.
There is no physical interpretation for new summation index
after the Poisson resummation.
The reason why they got the finite answer is the same as the one
of infinitely thin brane limit, that the anisotropic treatment
of the four momentum and the Kaluza-Klein mass.
We can apply the cutoff regularization in this thick brane setup
with the cutoff $\Lambda_c \sim \Lambda_s$ and get the same divergent
feature as long as the KK modes are truncated at the same scale
with the four dimensional momentum. If the brane is thicker
than the size of the fundamental scale, $\Lambda_s \ll \Lambda_c$,
then the answer may give UV insensitive result. However, 
the Kaluza-Klein spectrum and wave functions should be obtained
with considering these nontrivial brane configuration
and the simple Fourier decomposition can not be a solution any longer.
Certainly this is not a setup that was originally proposed.

In \cite{N} (and also in \cite{AHNSW}) the one loop effective potential
takes the expression that is completely dominated by the compactification
scale.
This reminds us of the equation \ref{eq:n}.
\bea \frac{1}{2} \frac{1}{\sinh \pi x} & = &
\sum_{n=-\infty}^{\infty} e^{-(2n+1)\pi x} \\ (x > 0) \nn
\eea
In the one loop correction to the Higgs mass,
after the sum of the infinite Kaluza-Klein modes,
we obtain the inverse of the hyperbolic sin function,
and it can be expanded as a power series of $e^{-2x}$.
However, at this moment there is no physical meaning
to this summation index $n$.
One thing that should be clearly distinguished is the physical meaning
of the four dimensional momentum and the Kaluza-Klein modes.
In the calculations using "Kaluza-Klein regularization",
the suppressed contributions are not for the KK modes
but for the four momentum.
Therefore, this result is nothing to do with the explanation
that the suppression is due to the separation of supersymmetry breaking
away from the brane since the relevant object feeling the fifth dimension is
the momentum along the fifth dimension, i.e., the Kaluza-Klein modes.
Furthermore, as already shown in the previous section,
supersymmetry is broken even in the bulk.
Therefore, the UV softness of the setup is an artifact of the infinite
Kaluza-Klein mode sum.

\section{summary and discussion}

We have shown that Scherk-Schwarz mechanism for the supersymmetry breaking
has a hard breaking nature such that radiative corrections to the Higgs
mass is quadratically divergent.
This is clearly seen from the contributions of each modes
with the isotropic momentum cutoff regularization.
The isotropic momentum cutoff regularization, 
applying the same cutoff scale 
to the four momentum and the Kaluza-Klein mass,
is the physical one since these are components
of the five momentum to which we should apply the cutoff.
The ``Kaluza-Klein regularization'' is an extremely anisotropic
regularization in this point of view.
The supersymmetry is broken everywhere except the fixed point
in the Scherk-Schwarz breaking,
and the exponential suppression of higher momentum (KK modes)
does not happen.
The apparent exponential suppression of higher four momentum
is nothing but a sequel of the infinite KK modes sum
and is not a reflection of the physics behind it.
We have shown that in the physical cutoff procedure,
the KK modes near the cutoff is not exponentially suppressed
and this entangles the four momentum and the result remains
very sensitive to the Kaluza-Klein modes near the cutoff.
As a result, quadratic divergences for the four momentum 
do not disappear.

In most calculations people use the infinite sum of KK modes.
However, it is doubtful since large log term is
unavoidable if we have different mass scales in a wide range especially
in the dimensional regularization scheme. Thus it is more
appropriate to use an effective theory after integrating out the
heavy particles. 
The best way to get the correction is the approach based on
integrating out and matching. In this way we can reproduce the
well known result of power law running of gauge couplings and
Yukawa couplings which reflect properties of higher dimensional
theories.
However, on the contrary to the usual theory in which the mass
spectrums are given below the fundamental scale
and we can integrate out heavy fields step by step,
the Kaluza-Klein decomposition always put their spectrums
across the fundamental scale of the higher dimensional setup.
It remains an open question whether we can find the correct
procedure dealing with the Kaluza-Klein modes
without knowing the quantum theory of gravity in higher dimensions.
There are interesting works done recently related to this
\cite{ACG,HPW,HPW2,ACG2,HPW3} utilizing the transverse lattice

\section*{acknowledgments}
The author thanks K. Choi, J. E Kim, P. Ko, T. Rizzo and M.
Schmaltz for discussions.

\end{document}